\documentstyle[referee]{mn}
\input epsf
\begin{document}
\title{The origin of peculiar jet-torus structure in the Crab nebula}
\author[S.S.Komissarov, Y.E.Lyubarsky]{S.S.Komissarov$^1$, Y.E.Lyubarsky$^2$\\
$^1$ Department of Applied Mathematics, the University of Leeds,
UK; e-mail: serguei@amsta.leeds.ac.uk\\ $^2$Physics Department,
Ben-Gurion University, P.O.B. 653, Beer-Sheva 84105, Israel;
e-mail: lyub@bgumail.bgu.ac.il}
\date{Received/Accepted}
\maketitle

\begin{abstract}
Recent discoveries of the intriguing ``jet-torus'' structure in the 
Crab Nebula and other pulsar nebulae prompted calls for re-examining 
of their theory. The most radical proposals involve abolishing
of the MHD approximation altogether and developing of purely
electromagnetic models. However, the classical MHD models of 
the Crab Nebula were hampered by the assumption of spherical 
symmetry made in order to render the flow equations easily 
integrable.  The impressive progress in computational relativistic 
magnetohydrodynamics in recent years has made it possible to study the 
Crab nebula via numerical simulations without making such a
drastic simplification of the problem. In this letter we present
the results of the first study of such kind. They show that the 
jet-torus pattern can be explained within MHD approximation when 
anisotropy of pulsar winds is taken into account. They also  
indicate that the flow in the nebula is likely to be much more 
intricate than it has been widely believed.
\end{abstract}
\begin{keywords}
pulsars:general -- supernova remnants -- ISM:individual:the Crab
Nebula -- ISM:jets and outflows -- MHD --- shock waves
\end{keywords}

\section{Introduction}
The Crab Nebula is a prototype compact synchrotron nebula
continuously powered by ultra-relativistic, magnetized wind
from young rapidly rotating pulsar. As this nebula is confined
within a nonrelativistic surrounding, the wind must terminate at a
shock wave. It is the wind plasma heated to relativistic
temperatures at this shock that fills the nebula and produces the
observed nonthermal electromagnetic emission from the radio to the
gamma-ray band (Rees \& Gunn 1974, Kennel \& Coroniti 1984).  
Indeed, early observations of the Crab Nebula 
revealed an apparent central hole in the nebula brightness distribution 
(Scargle 1969), which was identified with the termination shock in the 
theoretical models of the nebula. However, the assumption of spherical 
symmetry utilized in these models in order to simplify calculations was
shattered when recent X-ray and optical observations (Hester et
al. 1995, 2002; Weisskopf et al. 2000) revealed the inner
structure of the nebula in much greater details. They discovered
two jets aligned with the rotational axis of the pulsar and a
luminous torus or disc in its equatorial plane. As similar
structures have been found in other pulsar wind nebulae (Gaensler,
Pivovaroff \& Garmire 2001; Helfand, Gotthelf \& Halpern 2001;
Pavlov et al. 2001; Gaensler et al. 2002; Lu et al. 2002), we must
be dealing with quite generic a phenomenon.

The interaction between a relativistic MHD wind from the pulsar
and a dense, nonrelativistic surrounding is a much more
challenging problem in the case of anisotropic wind and there is
no much hope to find its analytical solution. Fortunately, recent
progress in numerical methods for relativistic gas dynamics and
MHD (Marti \& Muller 1999; Komissarov 1999) has made possible to
approach this problem numerically. The main results of the first 
attempt of such a study are briefly described in this letter. 
Full details will be presented elsewhere. 

\section{Structure of the pulsar wind in the far zone}
Although a self-consistent solution to the problem of pulsar wind
remains to be found, it is commonly accepted that far away from
the pulsar the wind can be considered as almost radial
supermagnetosonic outflow with purely azimuthal magnetic field and
anisotropic angular distribution of the energy flux (Michel 1982;
Beskin, Kuznetsova \& Rafikov 1998; Chiueh, Li \&Begelman 1998;
Bogovalov \& Tsinganos 1999). According to the simplified
split-monopole models of pulsar magnetospheres (Michel 1973;
Bogovalov 1999), the total energy flux density of the wind,
$f_{w}$, satisfies the following equation

\begin{equation}
f_{w}=\frac{f_0}{r^2}(\sin^2\theta+1/\sigma_0), 
\label{eflux}
\end{equation}
where $r$ and $\theta$ are the spherical coordinates whose polar
axis is aligned with the rotation axis of the pulsar. The first
term in the brackets represents the Poynting flux, whereas the second
one accounts for the small initial contribution of particles, 
$\sigma_0\gg 1$.
%This equation tells us that energy flows predominantly
%in the equatorial plane which is partly due to the fact that the
%azimuthal magnetic field of the wind vanishes along the rotational axis.
It was shown recently that the termination shock of such a wind is
not at all spherical. In fact, it is located significantly closer to the
pulsar along its rotational axis than in the equatorial plane
(Lyubarsky 2002; Bogovalov \& Khangoulyan 2002).
%Undoubtedly, this property alone makes a very
%convincing argument in favor of the anisotropic wind model for the Crab torus.

Crab's jets, as well as jets of other pulsars, appear to originate
from the pulsar (Weisskopf et al. 2000; Helfand, Gotthelf \&
Halpern 2001; Pavlov et al. 2001; 2003; Gaensler et al. 2002;
Hester et al. 2002;). This seems to indicate that they are formed
within the pulsar wind and the collimation by magnetic hoop stress
suggests itself. However, a closer look reveals a number of
problems with this explanation. First of all, such a collimation
is found to be extremely ineffective in ultra-relativistic flows
\cite{Bes,Chi,Bog-Tsi,Lyu-Eic}. Moreover, the direct observations of
proper motions in Crab's and Vela's jets indicate rather moderate
velocities of only $0.3\div 0.7c$ (Hester et al. 2002; Pavlov et
al. 2003). In order to overcome these problems, Lyubarsky (2002)
proposed that the jets are formed downstream of the termination
shock where velocities are no longer ultra-relativistic and the
magnetic collimation is much more effective. Because the
termination shock is much closer to the pulsar along its
rotational axis, this could give an impression of jets being
produced by the pulsar itself.

Properties of the pulsar wind nebula strongly depend on the wind
magnetization.
%, $\sigma$, defined as the ratio of the Poynting flux to
%the kinetic energy flux of the wind.
It is widely accepted that a typical pulsar wind is launched as a
Poynting-dominated flow but most of the electromagnetic energy is
transferred to particle along the way to the termination shock
(see, however, Begelman 1998). Although the problem of energy
conversion remains a subject of intensive debate (e.g. Melatos
2002), a number of mechanisms have been proposed in recent years.
Since the pulsar magnetic axis is inclined with respect to its
rotational axis, a significant fraction of the Poynting flux is
carried out by the component of electromagnetic field oscillating
with the rotational period of the pulsar. 
These small-scale waves can decay via various dissipation 
processes (Lyubarsky \& Kirk 2001; Melatos
2002; Lyubarsky 2003a; Kirk \& Skj{\ae}raasen 2003). Even if 
the dissipation time-scale is larger than the time of travel from 
the pulsar to the termination shock, these waves rapidly   
decay at the termination shock. In this case, the post-shock parameters 
are still the same as if the alternating fields had already annihilated 
in the upstream flow (Lyubarsky 2003b). For this reasons, we  
assume in our model of the pulsar wind that all these waves have already 
decayed and transferred their energy to particles. 
% In the equatorial plane, where most of
%the wind energy is transported, the magnetic field alternates
%changing sign every half-period.
%Such a flow is known as the striped wind$^{[15,16]}$.
%The energy of the alternating field dissipates because of the
%current starvation in current sheets separating strips with
%opposite magnetic field$^{[15,24,25]}$.
The rest of the Poynting flux is transported by the large-scale
azimuthal magnetic field. The exact latitudinal distribution of
this field remains to be found but it must vanish both along the
rotational axis, as any axisymmetrical magnetic field does, and in
the equatorial plane, because the average magnetic field in the
obliquely rotating magnetosphere is zero at the equator. We take
the residual magnetic field in the wind in the form
\begin{equation}
B=\sqrt{\frac{4\pi f_0}c}\,\frac{\xi}{r}\,
   \sin\theta\left(1-\frac{2\theta}{\pi}\right). 
\label{Bwind}
\end{equation}
The free parameter $\xi\le 1$ controls the wind magnetization.

\section{Numerical simulations}

The initial solution includes 1) a spherically symmetric shell of
cold dense gas expanding radially with velocity of 5000 km/s and
2) a radial ultra-relativistic wind with the energy flux density 
given by equation (1) with $\sigma_0=100$, and the magnetic field
described by equation (2). Provided the wind is ultra-relativistic, the
solution is not sensitive to the exact value of the wind Lorentz
factor, which we set to $\gamma_w=10$. To carry out these
axisymmetric simulations we used the Godunov-type scheme for
relativistic magnetohydrodynamics constructed recently by
Komissarov (1999).

The results for $\xi=0.3$ are presented in Figure 1 (overall
structure of the flow) and Figures 2, 3 (its central part). One
can see that instead of a single termination shock, the numerical
solution displays a whole complex of shocks. The equatorial,
weakly magnetized part of the pulsar wind terminates at almost
cylindrical shock crossing the equator, ``the Mach belt''. At
higher latitudes, the flow passes first through a highly oblique
"arch shock". Downstream of this shock the flow remains
supermagnetosonic with the typical velocity of $0.8\div 0.9c$.
Then it passes through another shock, ``the rim shock'', that
originates from the edge of the Mach belt. In contrast to the
spherically symmetric expansion assumed in the current theories of
pulsar nebulae, most of the downstream flow is confined to the
equatorial plane. Its typical velocity, $v \approx 0.6c$, agrees
with the measurements of proper motions in Crab's torus (Hester et
al. 2002). This equatorial outflow is eventually pushed back by
the pressure force and the magnetic hoop stress and forms a
backflow.

\begin{figure}
\leavevmode \epsffile[43 136 331 421]{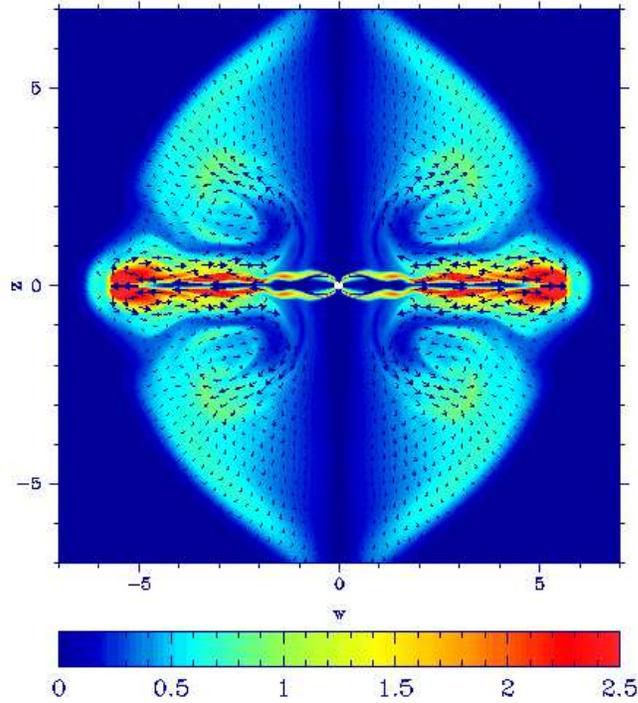} \caption{Fig.1. The
ratio of magnetic pressure to the gas pressure and velocity field
for the model with the magnetization parameter $\xi=0.3$. The
equatorial outflow is seen as the region of a particularly high
magnetic pressure.}
\end{figure}

\begin{figure}
\epsffile[43 136 331 421]{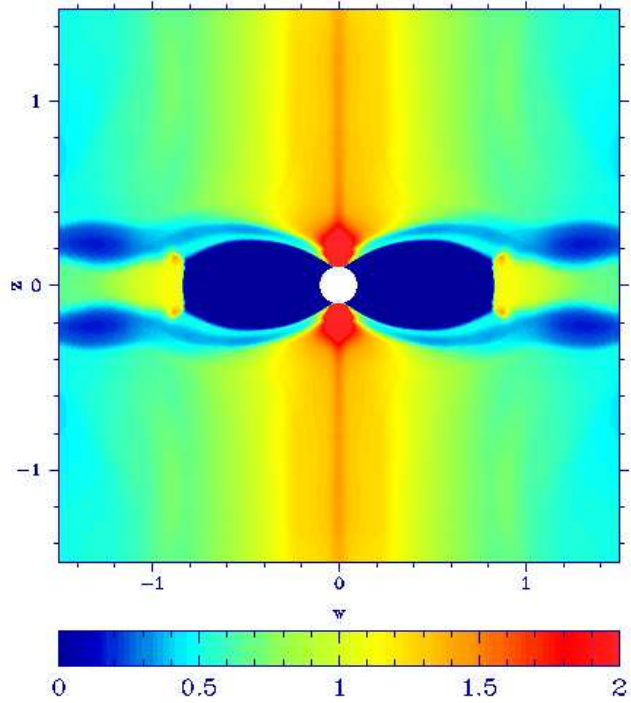} \caption{The gas pressure
distribution in the central part of the solution shown in figure
1. The white circle in the centre shows the inner boundary of the
computational domain whereas the adjacent region of dark blue
color shows the pulsar wind zone. The ``Mach belt'' shock runs
across the equatorial plane at the distance of $\approx 0.8$ from
the symmetry axis; the ``arch shock'' is seen as the upper and the
lower boundaries of the wind zone; the ``rim shocks'' originate
from the points of intersection of the Mach belt and the arch
shock.}
\end{figure}

\begin{figure}
\epsffile[43 136 331 421]{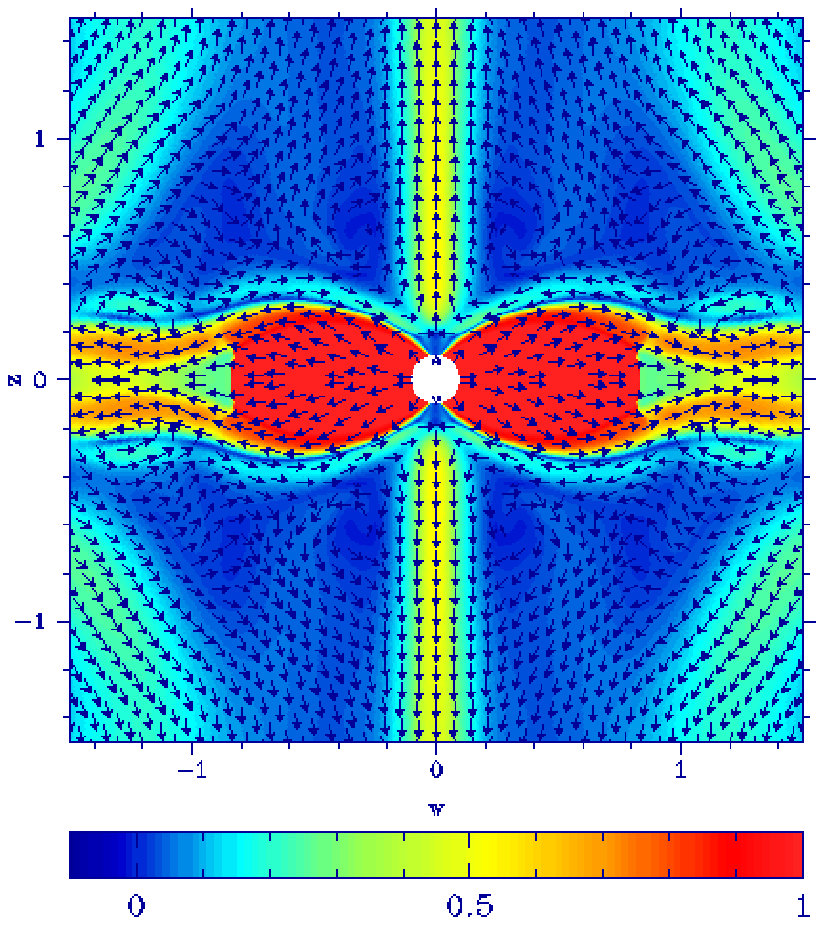} \caption{The velocity magnitude,
represented in color, and the flow direction, represented by
arrows, in the central part of the solution shown in figure 1.
Just above the equatorial outflow there seen a layer of backflow
converging toward the symmetry axis. This backflow provides plasma
for two transonic jets propagating in the vertical direction.}
\end{figure}
This backflow reaches the central region of the nebula and forms
large-scale vortices at intermediate latitudes. 
 
Another backflow is launched from the surface of the equatorial 
disc as the result of strong magnetic braking. This inner backflow 
eventually forms magnetically collimated polar jets.
%As we have already pointed out the magnetization of the pulsar
%wind is maximum near its polar axis. As the flow passes through
%the reconfinement shock and moves away from the axis, the magnetic
%field grows even further. At some point its hoop stress halts the
%outflow and pushes plasma back toward the axis. This leads to a
%significant increase of pressure near the cusp region of the
%reconfinement shock that drives the outflow along the polar axis.
%Thus, our results strongly support the idea that the Crab jets
%could indeed be formed downstream of the wind termination
%shock$^{[16]}$.
The typical velocity of these jets is $v_j \approx 0.5c$, which is
well in agreement with observed velocities of Crab's and Vela's
jets (Hester et al. 2002; Pavlov et al. 2003). Similar jets are
produced in the model with $\xi=0.5$ but not found in the solution
with $\xi=0.2$. The pulsar wind magnetization is traditionally measured 
by the parameter $\sigma$, defined as the ratio of the Poynting flux
to the kinetic energy flux of the wind (Kennel \& Coroniti 1984). 
It turns out that in our model the mean $\sigma\approx 0.1\xi^2$ 
Thus, the polar jets are produced for $\sigma \geq 0.01$, which is 
somewhat higher than the previous estimates  for the 
Crab Nebula based on simplified theoretical models
(Kennel \& Coroniti 1984;
Emmering \& Chevalier 1987; Begelman \& Li 1992).

Figure 4 shows the simulated synchrotron X-ray emission from the
inner part of the nebula obtained for the same orientation of the
wind relative to the observer as in the Crab nebula. In order to
create this image, we assumed that electrons (and positrons) are
injected at the termination shock and then suffer synchrotron
energy losses at a rate determined by the typical value of
magnetic field in the numerical solution. Because the magnetic
field in our simulations is purely azimuthal, it vanishes on the
symmetry axis and so does the synchrotron emissivity. In a real
jet, the strong velocity shear would generate the poloidal
component of the magnetic field, which does not have to vanish on
the axis. To take this into account, we added, only inside the
jet, the poloidal field aligned with the flow velocity at the
level of 30\% of the gas pressure.
%This made the jets more prominent
%compared to what they would be otherwise.
\begin{figure}
\epsffile[43 136 331 421]{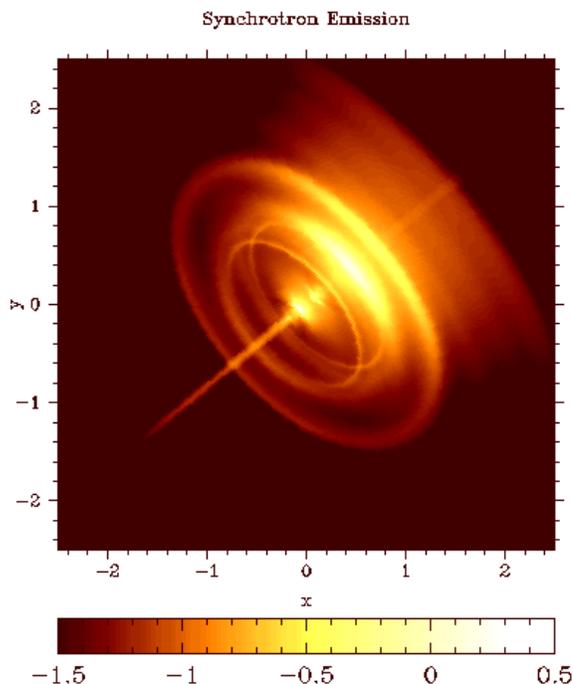}
\caption{Synchrotron X-ray image
for the solution shown in figures 1-3. The nebula is tilted to the
plane of the sky at the angle of $30^o$, just like the Crab
Nebula. The brightness distribution is shown in the logarithmic
scale.}
\end{figure}

\section{Discussion}
The obtained image is excitingly similar to the X-ray image of the
Crab nebula (Weisskopf et al. 2000). In addition to the polar
jets, one can see a system of rings which makes an impression of a
disc-like or even a toroidal structure. Well in agreement with the
observations these rings are brighter on the counter-jet side,
which is entirely due to the relativistic beaming effect. The
bright spot within the inner ring may correspond to the so-called
``sprite'' of the Crab nebula (Hester et al. 2002). This emission
originates in the high velocity plasma flowing just above the arch
shock where its velocity vector points directly toward the
observer.

On the whole, our model captures the main properties of the Crab
nebula remarkably well. There are, however, some qualitative and
quantitative differences which demand further investigation.
Crab's jet is not so straight and well collimated and it cannot be
traced that far away from the pulsar. It bends, spreads, and
eventually merges into the surrounding plasma (Weisskopf et al.
2000). The most likely reason for such a behavior is the
development of the kink instability (Begelman 1998), which is
suppressed in our simulations by the condition of axisymmetry.
Full 3-dimensional simulations are needed to overcome this
restriction.

The brightness contrast between the jet side and the counter-jet
side of the rings in the simulated maps is too high. Since this
asymmetry is entirely due to the relativistic beaming, it strongly
depends on the velocity field very close to the termination shock
and hardly depends on anything else. The main factor determining
this velocity field is the angular structure of the pulsar wind.
Thus, the brightness asymmetry of Crab's ``torus''
imposes strong observational constraints on the pulsar wind
models. Future simulations will be used to determine the model
parameters providing the best fit to the observational data.


\begin{thebibliography}{}
\bibitem[\protect\citename{Begelman }1998]{Beg}
   Begelman M.C., 1998, ApJ, 493, 291

\bibitem[\protect\citename{Begelman }1992]{Beg-Li}
   Begelman M.C., Li Z.-Y., 1992, ApJ, 397, 187

\bibitem[\protect\citename{Beskin et al. }1998]{Bes}
   Beskin V.S., Kuznetsova I.V., Rafikov R.R., 1998, MNRAS, 299,341

\bibitem[\protect\citename{Bogovalov }1999]{Bog99}
   Bogovalov S.V., 1999, A\&A, 349, 1017

%\bibitem[\protect\citename{Bogovalov }2001]{Bog01}
%   Bogovalov S.V., 2001, A\&A, 367, 159

%\bibitem{} Bogovalov, S. V., Aharonian, F. A., 2000, MNRAS, 313, 504

%\bibitem[\protect\citename{Bogovalov \& Khangoulyan }2002a]{Bog-Kha02a}
 %  Bogovalov S.V., Khangoulyan D.V., 2002a, Astron.Lett, 28, 373

\bibitem[\protect\citename{Bogovalov \& Khangoulyan }2002]{Bog-Kha02b}
   Bogovalov S.V., Khangoulyan D.V., 2002, MNRAS, 336, L53

\bibitem[\protect\citename{Bogovalov \& Tsinganos }1999]{Bog-Tsi}
   Bogovalov S.V., Tsinganos K., 1999, MNRAS, 305, 211

%   \bibitem[\protect\citename{Bucciantini et al.}2003]{Buc}
%Bucciantini N., Blondin J.M., Del Zanna L., Amato E., 2003, astro-ph/0303491

%\bibitem{} Chevalier R.A., 2002, in Neutron Stars in Supernova Remnants, ASP Conf. Ser., V. 271,
%Eds. P.O. Slane and B.M. Gaensler. San Francisco: ASP, p.125

\bibitem[\protect\citename{Chiueh \& Begelman }1998]{Chi}
   Chiueh T., Li Z.-Y., Begelman M.C., 1998, ApJ, 505, 835

%\bibitem[\protect\citename{Coroniti }1990]{Cor} Coroniti F.V., 1990, ApJ, 349, 538

\bibitem[\protect\citename{Emmering \& Chevalier }1987]{Emm-Che}
   Emmering R.T., Chevalier R.A., 1987, ApJ, 321, 334

\bibitem[\protect\citename{Gaensler et al. }2001]{Gae01}
   Gaensler B. M., Pivovaroff M. J., Garmire G. P., 2001, ApJ, 556, L107

\bibitem[\protect\citename{Gaensler et al. }2002]{Gae02}
   Gaensler B.M., Arons J., Kaspi V.M., Pivovaroff M.J., Kawai N.,
   Tamura K., 2002, ApJ, 569, 878

%\bibitem[\protect\citename{Goldreich \& Julian }1969]{Gol-Jul}
% Goldreich P. and Julian W.H., 1969, Ap.J., 157, 869.

\bibitem[\protect\citename{Helfand et al.}2001]{Hel}
   Helfand D.J., Gotthelf E.V., Halpern J.P., 2001, ApJ, 556, 380

\bibitem[\protect\citename{Hester et al. }1995]{Hes}
   Hester J.J. et al., 1995, ApJ 448, 240

\bibitem[\protect\citename{Hester et al. }2002]{Hes02}
   Hester J.J. et al.,
%Mori K., Burrows D., Gallagher J.S., Graham J.R., Halverson M.,
%Kader A., Michel F.C., and Scowen P.,
  2002, ApJ 577, L49

\bibitem[\protect\citename{Kennel \& Coroniti }1984]{Ken-Cor}
   Kennel  C.F., Coroniti F.V., 1984, ApJ, 283, 694

%\bibitem{} Kirk J. G., Ball L., Skj{\ae}raasen O., 1999, Astroparticle Phys., 10, 31

\bibitem[\protect\citename{Kirk \& Skj{\ae}raasen}2003]{Kirk}
   Kirk J.G., Skj{\ae}raasen O., 2003, ApJ, in press

\bibitem[\protect\citename{Komissarov }1999a]{Kom99a}
   Komissarov S.S., 1999a, MNRAS, 303, 343.

%\bibitem[\protect\citename{Komissarov }1999b]{Kom99b}
 %  Komissarov S.S., 1999b, MNRAS, 308, 1069

%\bibitem[\protect\citename{Kundt \& Krotscheck }1980]{Kun}
%   Kundt W., Krotscheck E., 1980, A\&A, 83, 1

%\bibitem[\protect\citename{Ingraham}1973]{Ing} Ingraham, R. L.,  1973, ApJ, 186, 625

\bibitem[\protect\citename{Lu et al. }2002]{Lu}
   Lu F.J., Wang Q.D., Aschenbach B., Durouchoux P., Song
   L.M.,  2002, ApJ, 568, L49

\bibitem{} Lyubarsky Y.E., 2002, MNRAS, 329, L34

\bibitem[\protect\citename{Lyubarsky }2003a]{Lub03a}
   Lyubarsky Y.E., 2003a, MNRAS, 339, 765

\bibitem[\protect\citename{Lyubarsky }2003b]{Lub03b}
   Lyubarsky Y.E., 2003b, MNRAS, submitted

\bibitem[\protect\citename{Lyubarsky \& Eichler }2001]{Lyu-Eic}
   Lyubarsky Y.E., Eichler D., 2001, ApJ, 562, 494

\bibitem[\protect\citename{Lyubarsky \& Kirk }2001]{Lub01}
   Lyubarsky Y.E., Kirk J.G., 2001, ApJ, 547, 437
%\bibitem{} Heyvaerts, J., 1996, In: Plasma Astrophysics, Chiuderi
%C., Einaudi G. (eds.), Springer, 31

\bibitem{} Marti J.M. Muller E., 1999, electronic journal {\it Living Reviews in
Relativity} at
http://www.livingreviews.org/Articles/Volume2/1999-3marti/index.html
%22%

\bibitem{} Melatos, A. 2002,
 in: P.O.Slane and B.M.Gaensler (eds.), Neutron Stars in Supernova Remnants,
 ASP Conference Series, Vol. 271, San Francisco: ASP, p.115

\bibitem[\protect\citename{Michel }1973]{Mic73}
   Michel F.C., 1973, ApJ, 180, 133

%\bibitem[\protect\citename{Michel }1974]{Mic74} Michel F.C., 1974, ApJ, 187, 585

\bibitem[\protect\citename{Michel }1982]{Mic82}
   Michel F.C., 1982, Rev.Mod.Phys., 54, 1

\bibitem[\protect\citename{Pavlov et al. }2001]{Pav}
   Pavlov G.G., Kargaltsev O.Y.; Sanwal D.; Garmire G.P.,
   2001 ApJ 554, L189

\bibitem[\protect\citename{Pavlov et al. }2003]{Pav03}
   Pavlov G.G., Teter M.A., Kargaltsev O.Y.; Sanwal D., 2003, ApJ,
   in press; astro-ph/0305510

\bibitem[\protect\citename{Rees \& Gunn }1974]{Ree-Gun}
   Rees M.J., Gunn J.E., 1974, MNRAS 167, 1

\bibitem[\protect\citename{}1969]{}
   Scargle J.D., 1969, ApJ, 156, 401
%\bibitem[\protect\citename{Tomimatsu }1994]{Tom} Tomimatsu A., 1994, PASJ, 46, 23

%\bibitem[\protect\citename{Usov }1975]{Uso} Usov V.V., 1975, ApSS, 32, 375

% \bibitem[\protect\citename{van der Swalluw}2003]{Swa}
%  van der Swalluw E., 2003, astro-ph/0303661

%\bibitem[\protect\citename{Weiler \& Panagia }]{Wei-Pan}
 %  Weiler K.W., Panagia N., 1978, A\& A, 70, 419

\bibitem[\protect\citename{Weisskopf C. et al. }2000]{Wei}
   Weisskopf C. et al., 2000, ApJ, 536, L81

\end{thebibliography}
\end{document}